\documentclass[12pt,preprint]{aastex}

\newcommand{\sgra}{Sgr A$\textrm{*}$}

\renewcommand {\deg}   {\mbox{$^\circ$}}

\newcommand   {\arcs}  {\mbox{$^{\prime\prime}$}}

\newcommand   {\kms}   {\mbox{km\,s$^{-1}$}}
\renewcommand {\ga}    {\mbox{\rlap{\hbox{\lower5pt\hbox{$\sim$}}}\hbox{$>$}}}
\renewcommand {\la}    {\mbox{\rlap{\hbox{\lower5pt\hbox{$\sim$}}}\hbox{$<$}}}

\begin{document}
\pagenumbering{arabic} 
\def\kms {\hbox{km{\hskip0.1em}s$^{-1}$}} 
\voffset=-0.8in

\def\msol{\hbox{$\hbox{M}_\odot$}}
\def\lsol{\hbox{$\hbox{L}_\odot$}}
\def\kms{km s$^{-1}$}
\def\Blos{B$_{\rm los}$}
\def\etal   {{\it et al.}}                     
\def\psec           {$.\negthinspace^{s}$}
\def\pasec          {$.\negthinspace^{\prime\prime}$}
\def\pdeg           {$.\kern-.25em ^{^\circ}$}
\def\degree{\ifmmode{^\circ} \else{$^\circ$}\fi}
\def\ut #1 #2 { \, \textrm{#1}^{#2}} 
\def\u #1 { \, \textrm{#1}}          
\def\nH {n_\mathrm{H}}
\def\ddeg   {\hbox{$.\!\!^\circ$}}              
\def\deg    {$^{\circ}$}                        
\def\le     {$\leq$}                            
\def\sec    {$^{\rm s}$}                        
\def\msol   {\hbox{$M_\odot$}}                  
\def\i      {\hbox{\it I}}                      
\def\v      {\hbox{\it V}}                      
\def\dasec  {\hbox{$.\!\!^{\prime\prime}$}}     
\def\asec   {$^{\prime\prime}$}                 
\def\dasec  {\hbox{$.\!\!^{\prime\prime}$}}     
\def\dsec   {\hbox{$.\!\!^{\rm s}$}}            
\def\min    {$^{\rm m}$}                        
\def\hour   {$^{\rm h}$}                        
\def\amin   {$^{\prime}$}                       
\def\lsol{\, \hbox{$\hbox{L}_\odot$}}
\def\sec    {$^{\rm s}$}                        
\def\etal   {{\it et al.}}                     
\def\la{\lower.4ex\hbox{$\;\buildrel <\over{\scriptstyle\sim}\;$}}
\def\ga{\lower.4ex\hbox{$\;\buildrel >\over{\scriptstyle\sim}\;$}}
\def\refitem{\par\noindent\hangindent\parindent}
\oddsidemargin = 0pt \topmargin = 0pt \hoffset = 0mm \voffset = -17mm
\textwidth = 160mm  \textheight = 244mm
\parindent 0pt
\parskip 5pt

\shorttitle{Sgr A*}
\shortauthors{}

\title{ALMA and VLA Observations: Evidence for Ongoing\\ 
Low-mass Star Formation  near Sgr A*}
\author{F. Yusef-Zadeh$^1$, W. Cotton$^2$, M. Wardle$^3$, M. J. Royster$^1$,\\ 
D. Kunneriath$^2$,  D. A. Roberts$^1$, A. Wootten$^2$, \& R. Sch\"odel$^4$ 
}
\affil{$^1$Department of Physics and Astronomy and CIERA, Northwestern University, Evanston, IL 60208}
\affil{$^2$National Radio Astronomy Observatory,  Charlottesville, VA 22903}
\affil{$^3$Department of Physics and Astronomy and 
Research Center for Astronomy, Astrophysics\\
and Astrophotonics, Macquarie University, Sydney NSW 2109, Australia}
\affil{$^4$Instituto de Astfisica de Andalucia (CSIC), 
Glorieta de la Astronomia S/N, 18008 Granada, Spain}

\begin{abstract}
Using the VLA, we recently detected a large number of 
protoplanetary disk (proplyd) candidates lying within a couple of light
years of the massive black hole Sgr A*.  The bow-shock appearance of
proplyd candidates point toward the young massive stars located near Sgr
A*.  Similar to Orion proplyds, the strong UV radiation from the cluster
of massive stars at the Galactic center is expected to photoevaporate
and photoionize the circumstellar disks around young, low mass stars,
thus allowing detection of the ionized outflows from the photoionized layer surrounding 
cool and dense gaseous disks. To confirm this picture, 
ALMA observations  detect millimeter  emission at 226 GHz from   five 
proplyd candidates 
that had been detected at 44 and 34 GHz with the VLA.  
We present  the derived disk masses for four sources  as a function of the assumed dust temperature.  
The mass of
protoplanetary disks  from cool dust emission ranges between 0.03 -- 0.05 \msol. 
These  estimates are consistent with the disk masses found in star forming sites in the Galaxy. 
These measurements show  the presence of on-going star formation with the implication that gas clouds 
can survive near Sgr A* and  the relative importance 
of high vs low-mass star formation in the strong tidal and
radiation fields of the Galactic center. 
\end{abstract}

\keywords{Galaxy: center - clouds - ISM: general - ISM - radio continuum - stars: protostars}

\section{Introduction}

The Galactic center hosts a population of young stars centered on 
a 4$\times10^6$ \msol\, black hole which coincides with  the strong radio source 
Sgr A* (Ghez et al.  2008; Gillessen et al. 2009).  
A stellar cluster of about one hundred young massive OB and WR stars 
lie within 1 and 10$''$ (0.04--0.4 pc) of Sgr A* 
(Paumard et al.  2006; Lu et 
al.  2009). An important  question regarding star formation near supermassive black holes (SMBHs)
is whether tidal shear in the vicinity of SMBHs is able to completely
suppress star formation or whether it induces disk-based star formation,
entirely distinct from the standard cloud-based mode observed in the Galactic
disk.  The stellar disk, the infrared excess sources as well as the molecular
ring orbiting Sgr A* in the inner few parsecs of the Galactic center are
excellent testing grounds to examine star formation in an extreme tidal
environment. The study of these sources near Sgr A* provides us with a
fantastic opportunity with far reaching implications for understanding star
formation in the nuclei of more active galaxies hosting truly supermassive
black holes.

A number of recent studies suggest that a  disk-based mode of star formation occurred 
between  4 to 8 million years ago 
within 0.5 pc of Sgr A* (e.g., Genzel et al.  2010). 
There are also several signatures of star formation beyond 
this region suggesting a cloud-based  mode of star formation (Geballe et al. 2006; Muzic et al. 2008; 
Eckart et al. 2013; Yusef-Zadeh et al. 2013, 2015a, 2016). 
If indeed  star formation took place near Sgr A*, 
this region  should contain numerous  
low mass stars with circumstellar disks (Haish, Lada \& Lada 2001).  

Low-mass stars at the Galactic center distance of 8 kpc are too faint, too far and highly extincted to be detected at 
near-IR and optical wavelengths. However, the strong UV radiation from the cluster of massive stars will 
photoevaporate and photoionize the circumstellar disks around young, low mass stars, thus allowing radio detection of 
ionized outflows (Johnstone et al. 1998; St\"orzer \& Hollenbach 1990). 
We have recently detected 44 candidate protoplanetary disks (proplyds) at 34 GHz with cometary morphology 
within 20$''$ of Sgr A* 
(Yusef-Zadeh et al. 2015a).  The short expansion time scale and
the low density of ionized gas associated with the cometary  structures provide strong arguments in favor of
proplyds (Li \& Loeb  2013; Yusef-Zadeh \etal\, 2015a).
Recent H42$\alpha$ recombination line observations indicate 
that these candidates have radial velocities ranging between 130 and 150 \kms\, (Tsuboi et al. 2016a), 
thus suggesting that 
they lie close to Sgr A*.  
Near-IR emission from two proplyd candidates indicate a layer of hot dust emission 
separated from the photoionized layer (Yusef-Zadeh et al. 2015b), thus implying 
that these candidates have disks with surface layers of hot dust and warm molecular gas 
that are  photo-ionized by the stellar cluster near Sgr A*.

ALMA observations presented here are motivated to  search for cool dust emission from  
circumstellar disk candidates and establish their proplyd nature with the implication that low-mass star formation 
is taking place near Sgr A*. 
Our measurements  confirm that the brightest proplyd sources detected at 34 GHz 
have mm counterparts.   We identify  five  proplyd candidates  that have mm counterparts 
with  disk mass estimates   similar to those 
found in the Orion Nebula and NGC 2024 (Mann et al. 2014), assuming the  dust temperature is 100K 
(Lau et al. 2013). 
We also determine the disk mass as a function of the dust temperature.



\section{Observations and Data Reduction}

ALMA   and 
the Karl G. Jansky Very Large Array       
(VLA)\footnote{Karl G. Jansky Very Large Array (VLA) of the National Radio
Astronomy Observatory is a facility of the National
Science Foundation, operated under a cooperative agreement by Associated Universities, Inc.}
 observations were carried out as part of a multi-wavelength observing campaign to monitor the 
flux variability of Sgr A*. A detailed account of these observations  will be given elsewhere. Here we focus on 
observations related to the cluster of proplyd candidates located about 20$''$ NE of Sgr A*. 
Observations 
were obtained  on  July 12 and July 18, 2016, as part of the director's discretionary 
time given to us to join the observing campaign. 

The ALMA 230 GHz data consisted of two spectral windows centered on
218.3 and 238.0 GHz, each 1.87 GHz wide.
Bandpass and delay calibration was based on J1924-2914.  Cross hand
gain calibration was based on Titan and Pallas which were assumed to
be unpolarized and subsequent calibration averaged the parallel hand
(XX and YY) data sets.
Initial amplitude and phase calibration was based on 1744-3116 with an
assumed flux density of 0.26 Jy at 234 GHz.
Phase self calibration followed by amplitude and phase calibration was
used to reveal the low level emission but adds uncertainty to the
overall amplitude gain calibration.
The editing and calibration of the data was carried
out by  OBIT (Cotton 2008) before all the spectral windows were averaged prior to constructing 
final images.  The July 18 data which 
has  a higher spatial resolution ($0.36''\times0.25''$) than the  July 12 data are presented here.  

We also used calibrated ALMA archival data at 100 GHz from Cycle 0 (project code 2011.0.00887.S) observed on May 18, 
2012 with nineteen 12-m antennas. Neptune and Titan were used for flux calibration, while J1924-292 and NRAO530 were 
the bandpass and phase calibrators, respectively. The 100 GHz dataset contains 
contained 4 spectral windows of 2 GHz bandwidth.
We imaged the continuum  by combining all four spectral windows after phase and amplitude 
self-calibrations, 
using CASA version 4.5.3. 
The final sensitivity in the 100 GHz image presented here is ~0.6 mJy/beam and the beam size is 
1.58"$\times$1.31", P.A.=-87.5$^\circ$.



Radio continuum observations were   carried out  with the VLA in its B configuration 
on the same days that ALMA observations took place.
We used Ka-band (8.7 mm) and Q-band (7mm) 
with the 3-bit sampler system, which provided full polarization correlations in 4 
basebands, each 2 GHz wide. 
Each baseband was composed of 16, 128 MHz wide,  
subbands.  Each subband was made up of 64 channels,  each   2 MHz wide.  
We used 3C286 to 
calibrate the flux density scale and used 3C286 and J1733-1304 (aka NRAO530) to calibrate the bandpass.  We used 
J1744-3116 to calibrate the complex gains.  
We  constructed   a Q-band  image of the 
30\arcs\ surrounding \sgra\, with a spatial resolution of 
$\sim0.4''\times0.2''$ (PA$=-1.6^\circ$).

\section{Results}

Figure 1a,b show 225 and 44 GHz grayscale images of the inner 35$''\times25''$ of the Galactic center. The 
similarity of the mini-spiral in radio, mm, submm and mid-IR bands suggests that the emission arises from layers of 
hot dust, 
free-free and cool dust (Viehmann et al. 2006; Eckart et al. 2013; Kunneriath et al. 2012; Tsuboi et al. 2016b; 
Yusef-Zadeh et al. 2016). 
The ionized features are clearly accompanied by dust and molecular gas that is photoionized by 100-200
 OB stars distributed within 10$''$ of Sgr A*. 
 Although, the separation of layers of dust and gas emission is complicated, it is clear that the interior 
to the 2-pc circumnuclear molecular ring is not entirely filled by ionized gas, as had been assumed in 
the past (see the review by Genzel et al. 
2010).  The new ALMA images of the mini-spiral 
provide  a  paradigm  shift supporting direct evidence for 
fuels needed to  accrete  onto Sgr A* and to form stars in the extreme environment of 
Sgr A*. 
Based on a recent ALMA study of the mini-spiral, Tsuboi et al. (2016b) estimate molecular clumps of 10-100 \msol\, 
distributed in the mini-spiral.

Figure 2a shows a  region associated with 
the NE arm of the mini-spiral (see the box in Figure 1) where proplyd candidates are detected. 
There are at least five 44 GHz continuum sources that have 226 GHz counterparts.  We also  detect 100 GHz 
emission from a concentration of proplyds but the spatial resolution is too poor to identify individual proplyd sources.
In spite of the 100 GHz low resolution, most of the proplyd candidates have 100 GHz counterparts.  
There are also 
other sites where proplyd candidates with mm counterparts are detected along the N and E arms of the mini-spiral. 
However, the ALMA spatial resolution at 100 GHz is not sufficient to identify individual proplyd 
candidates. Figure 2b,  displaying  a smaller region than Figure 2a,  
shows contours of 100 GHz emission which coincides with proplyd candidates detected at 34 GHz 
(Yusef-Zadeh et al. 2015a). 

Table 1 shows the parameters of Gaussian fits to five individual sources that 
are identified  as proplyd candidates 
(Yusef-Zadeh et al. 2015a).  Columns 1 to 5 give the source numbers, the coordinates and the peak  flux densities at 
44 and 226 GHz, respectively.  Column 6 calculates the dust emission by subtracting the 44 GHz flux densities from 
those of 226 GHz with the assumption that  the 44 GHz emission is optically thin and is not contaminated by dust emission. 
Columns 7 and 8  give the disk mass
and   the names of individual sources from Yusef-Zadeh et al. (2015a), respectively.

To estimate the disk masses, we assume that the 44 GHz fluxes are dominated by optically-thin bremsstrahlung with 
electron temperature 8000 K. The frequency-dependence of the gaunt factor (e.g., equation
 10.6 of Draine 2002) implies that 
the bremsstrahlung flux at 226 GHz is 0.76 times the 46 GHz flux. The balance of the 226\,GHz continuum is presumed to 
be thermal continuum emission from dust, and is given as $F_\mathrm{dust}$ in Table 1. We follow Mann et al's (2014, 
2015) approach to estimating proplyd masses in the Orion nebula and NGC 2024 clusters, in assuming that the emission 
is optically thin and is dominated by dust in the outer disk with some temperature $T_d$. Then 
\begin{equation}
    M_\mathrm{disk} = \frac{d^2 F_\mathrm{dust}}{\kappa_\nu B_\nu(T_d)} 
\end{equation}
(e.g. Beckwith et al.\ 1990) where $d=8$\,kpc is the distance to the Galactic center, $\kappa_\nu$ is the dust grain 
opacity at 0.856\,mm, and $B_\nu$ is the Planck function.  We set $\kappa_\nu = 
0.068\,\mathrm{cm}^{2}\,\mathrm{g}^{-1}$, twice that adopted by Mann et al. (2014)  and Beckwith et al (1990) to account for 
the twice solar metallicity at the Galactic center.  
Lau et al. (2013) derive a dust temperature at the location of proplyd
candidates ranging between 90 and 105K based on the 9 to 37 $\mu$m intensity ratios (see their Figure 8).  
We adopt $T_d = 100$\,K, representative of the inner parsec of the 
Galaxy, where dust is heated by the UV radiation from hot stars (Latvakoski et al.\ 1999; Lau et al. 2013).  
This is substantially 
higher than the $\sim 20$\,K adopted by Mann et al.\ (2014, 2015) that is characteristic of molecular clouds beyond 
10\,pc (Pierce-Price et al.\ 2000) or within the Galactic disk.

The  disk masses derived using equation 1
are listed in Table 1 for sources 1--4, and a 2-sigma upper limit is provided for source 5.  There is considerable 
uncertainty in these estimates. The 0.4$''$ beam at 226 GHz may contain emission from residual cloud material within 
1600 AU of the young star.  The estimated disk mass is inversely proportional to the adopted values of $\kappa_\nu$ 
and $T_d$.  
Figure 3 shows the derived disk masses for sources 1--4 as a function of the assumed dust temperature.  We 
see that for reasonable values of $T_d$ the estimates are consistent with the $\sim0.03-0.05\,\msol$ 
disk masses. 
The proplyd masses in Orion are indeed much lower (Mann et al 2014) , but these are relatively old disks (~1-2 Myr),
so that the masses have declined over time due to internal disk evolution and photo evaporation.  Even then Mann et
al. (2014) found that the proplyd disk masses in Orion range up to 0.078\, \msol. While many disks have
masses of order 0.001 \msol,  Mann et al. (2014) detected five sources that are more massive than 0.01\, \msol.
However, a comparison with proplyds in a young cluster is more appropriate.  Mann et al (2015) surveyed
NGC 2024 (age $\sim$0.5 Myr) and found a greater fraction of disks with masses exceeding 0.01 \msol,
 consistent with our estimated masses for T$_d\sim$100K.


These estimates are systematically low if the emission is not completely optically thin.  To explore this, consider a 
uniform disk with semi-major and minor axes $a$, and $b$ on the sky (in cm) that is emitting flux $F_\nu$ .  The 
optical depth through the source is
\begin{equation}
    \tau_\nu = -\ln \left(1-\frac{d^2F_\nu}{\pi abB_\nu}\right)\,,
\end{equation}
and its column density is $\tau_\nu / \kappa_nu$, so its mass is 
\begin{equation}
M_{\mathrm{disk}, \tau} =  \pi ab\, \tau_\nu/\kappa_\nu.    
\end{equation}
By way of example, 
we adopt $ab = (300\,\mathrm{AU})^2$ and compute the disk mass for source 4, which has the largest flux and for 
which optical depth effects will be most significant.  This is plotted as the blue dashed curve in Figure 3.
We see that optical depth effects add at most 10\% to the derived disk mass but are typically far smaller. In 
particular, the effect is small for the fainter sources.


\section{Discussion}

A key question is whether on-going star formation is taking place along the inner couple of  parsecs of the 
Galactic center. 
Here, we have shown the evidence for 
gaseous disks ionized by the external radiation 
 near Sgr A*, thus  providing  strong support for the formation of young stars under extreme condition.
After the comparison of  near-IR and radio images, the newly detected proplyds do not coincide with any 
known massive stars. So, these sources can not be due to dusty HII regions produced by 
massive  stars. Also, low mass stars at the Galactic center are too faint to be detected at near-IR wavelengths.  
The  upper limits to the 
infrared flux from the Galactic center proplyd candidates are consistent with gaseous disks 
orbiting low mass stars at a projected distance of $0.6-0.8$ pc from Sgr~A*.
A recent study indicates that despite the strong tidal and UV radiation fields at the Galactic 
center, the formation of low-mass stars is easier than high mass stars near the strong gravitational potential of 
Sgr A* (Wardle \& Yusef-Zadeh 2017). This is  because of the criteria for the collapse of a cloud, under Roch and 
Jeans limits,  depends on the distance from the massive black hole (Wardle and Yusef-Zadeh 2017). 
Thus the collapse of low-mass cloud is favored. 

Although the origin of massive stars within 0.5 pc of Sgr A* is discussed in the context of the
instability of massive gaseous disks and a cloud capture by Sgr A* (Nayakshin et al. 2007; Wardle and
Yusef-Zadeh 2008), the formation of isolated low mass star formation has not been fully understood.
Jalali et al. (2014) have recently suggested that the tidal compression of a clump of molecular gas
launched from the circumnuclear ring in a highly eccentric orbit can attain densities greater than the
Roche density. In another study, the gravitational stability of a cloud suggests that other forms of
external pressure such as shocks and/or  radiation, lower the Roche density, thus gravitation instability of
a cloud near Sgr A* is possible (Yusef-Zadeh and Wardle 2016; Wardle and Yusef-Zadeh 2017).

Embedded high mass stars are easily detected in the near to mid IR (Schoedel et al. 2009). 
In addition, there are
several constraints on embedded high-mass YSOs (Yusef-Zadeh et al. 2015b; Viehman et al. 2006). 
As for low mass stars, the $\sim$50\% completeness limit is around 18
magnitude, and they are likely to be sub-giants 
on the ascending branch. Solar mass main sequence  stars have K magnitude $\sim21$ 
at the Galactic center (see Alexander 2005).
The detection of low-mass stars would provide a strong constraint on the  IMF near a supermassive
black hole since arguments have been made that the Galactic center IMF 
may be  different than that  in  the Galactic disk (e.g., Bartko et al. 2010).
A top-heavy IMF has been suggested for stars in the central cluster
(Bonnell \& Rice 2008; Bartko et al. 2010; Lu et al. 2013). However, 
IMF determination is difficult in this region of the Galaxy and is  prone to systematic effects 
(see  Stolte et al. 2005; Kim et al. 2006; Hosek et al. 2015).

The morphology of individual sources and their clustering are the main discriminant to identify
protostellar candidates at radio.  The census of proplyds is incomplete due to extended emission and
confusing sources dominating the eastern and northern arms of the mini-spiral.  In spite of these
difficulties, we have recently found two additional clusters of proplyd candidates at the western edge
of the northern arm which will be reported in the future.  Our measurements suggest a total of about 100
sources that are identified in the regions where extended emission is not dominant.

Although we don't have an accurate number of low-mass stars within 
the inner pc of Sgr A*,  we estimate that there are more than  hundred  proplyd candidates detected at 
radio wavelengths. If we assume 100 massive stars  with $M > 10$ \msol, 
the expected   number of low mass-stars between 1--3 \msol\, 
with an IMF scaling as M$^{-\alpha}$ where $\alpha$=1.25 is 
consistent with observations. 
Columns 1 to 3 of Table 2 lists the value of  $\alpha$, the  number of stars 
with mass range between 0.5-1  and 1-3 \msol\, respectively.  One of the implication of 
the evidence for low mass star formation near Sgr A* 
is that it rules out an  IMF that is truncated near the bottom.  

In analogy with the Orion cluster, we examine the  disk mass of proplyds as a function 
of distance from the
the extreme UV-dominated region of the stellar cluster surrounding Sgr A*. 
The lack of massive disks may imply a rapid dissipation of disk masses 
near Sgr A*.  
Unlike the Orion proplyds, the Galactic center proplyds suffer not only from the strong UV radiation 
field but also from the tidal field of  Sgr A*. Tidal truncation will be significant for proplyds lying 
very close to Sgr A*.   Our limited sample does not allows us to examine the disk mass as a function of distance 
from Sgr A*.  However, 
the disk mass exceeds the minimum mass solar nebula (MMSN), $10^{-2}$ \msol\, thus the disk 
may be the birthplace of planetary systems near Sgr A*.  
Future sensitive ALMA observations will place strong constraints on 
the IMF near the Galactic center and provides 
further understanding of the role of 
environment on disk evolution and  the birth of planets.

Acknowledgments:                                          
This work is partially supported by the grant
AST-1517246 from the NSF. 
RS acknowledges funding from the European Research Council under 
the European Union's Seventh Framework Programme (FP7/2007-2013) / ERC grant agreement  [614922].
This paper makes use of the following ALMA data: 
2015.A.00021.S (Principal Investigator Gunther Witzel)
ALMA is a partnership of ESO (representing its member states), NSF (USA) and NINS
(Japan), together with NRC (Canada) and NSC and ASIAA (Taiwan), in cooperation with the
Republic of Chile. The Joint ALMA Observatory is operated by ESO, AUI/NRAO and NAOJ.


\vfill\eject

\begin{deluxetable}{cccccccc}
\tablecaption{Parameters of Gaussian fits to 44 and 226  GHz proplyd sources}
\tablecolumns{8}
\tabletypesize{\scriptsize}
\tablewidth{0pt}
\setlength{\tabcolsep}{0.04in}
\tablehead{
    \colhead{Proplyd} & \colhead{$\alpha$ (J2000)} & \colhead{$\delta$ (J2000)}
        & \colhead{F$_{\textrm{44.2 GHz}}^{\textrm{Peak}^a}$}
        & \colhead{F$_{\textrm{226  GHz}}^{\textrm{Peak}}$}
        & \colhead{F$_{\textrm{dust}}$}   & \colhead{M$_{\textrm{disk}^b}$}
        & \colhead{Notes} \\
    \colhead{Name}    & \colhead{$17^h45^m$ ($s$)} & \colhead{$-29^\circ00^\prime$ ($^{\prime\prime}$)}
        & \colhead{(mJy)}                                     & \colhead{(mJy)}
        & \colhead{(mJy)}                                     & \colhead{(\msol)}
        & \colhead{}
}
\startdata
1  & 41.0856 & 25.930 & 0.71$\pm$0.12 & 1.49$\pm$0.47 & 0.95$\pm0.48$ & 0.029$\pm0.015$      & P7$^c$ \\
2  & 41.1013 & 25.780 & 1.14$\pm$0.11 & 2.52$\pm$0.46 & 1.66$\pm0.47$ & 0.050$\pm0.014$    & P8$^b$  \\
3  & 41.2688 & 25.507 & 0.68$\pm$0.12 & 2.16$\pm$0.46 & 1.65$\pm0.47$ & 0.050$\pm$0.014  &  P28$^b$ \\
4 & 41.2691 & 24.659 & 0.76$\pm$0.12 & 1.62$\pm$0.47 & 1.05$\pm0.48$  & 0.032$\pm$0.014   &  P26$^b$ \\
5 & 41.2584 & 24.961 & 0.87$\pm$0.11 & 0.98$\pm$0.49 & 0.32$\pm0.50$  & $<$0.045  & P26$^b$ \\  
\enddata
\tablenotetext{a}{Convolved to the resolution of the 226 GHz image
($0.^{\prime\prime}42\times0.^{\prime\prime}35$, PA = $80^\circ$)}
\tablenotetext{b}{assuming dust temperature is 100K}
\tablenotetext{c}{Yusef-Zadeh et al. (2015a)}
\end{deluxetable}

\begin{deluxetable}{lcc}
\tablecaption{Expected number of low-mass stars extrapolated from  100 massive stars with M$>$10 \msol}
\tabletypesize{\scriptsize}
\tablecolumns{3}
\tablewidth{0pt}
\setlength{\tabcolsep}{0.04in}
\tablehead{
\colhead{$\alpha$} & \colhead{0.5-1 $\msol$} & \colhead{1-3 $\msol$}
}
\startdata
     0.5 &	   13.24 &	   18.73\\
    0.75 &	   30.17 &	   35.88\\
       1 &	   68.26 & 	   68.26\\
    1.25 &	   153.3 &	   128.9\\
     1.5 &	   341.9 &	   241.8\\
    1.75 &	   757.4 &	   450.3\\
       2 &	    1667 &	   833.3\\
    2.35 &	    4979 &	    1953\\
\enddata
\end{deluxetable}

\begin{figure}
\center
\includegraphics[scale=0.5,angle=0]{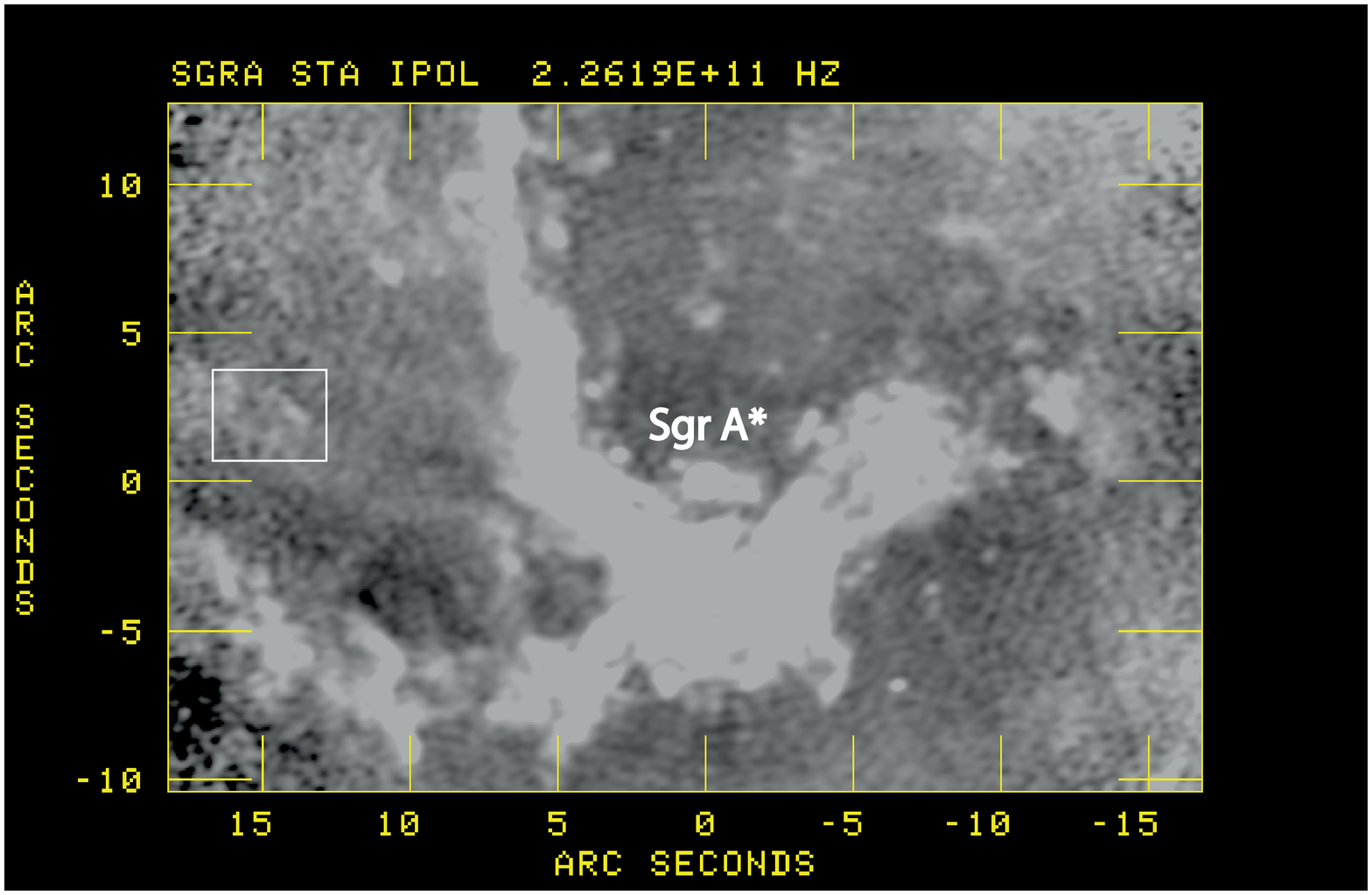}
\center
\includegraphics[scale=0.5,angle=0]{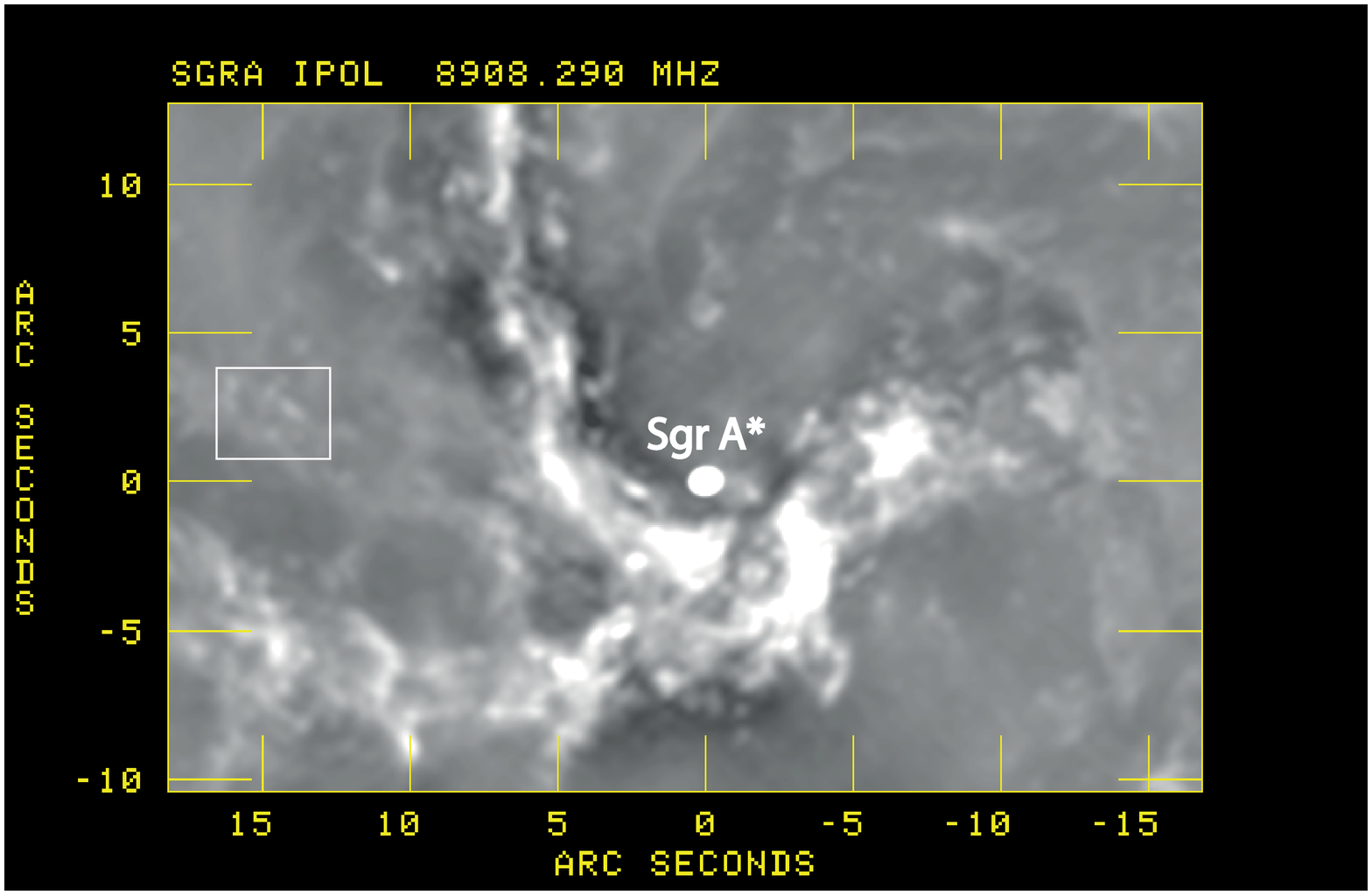}
\caption{
{\it (a) Top}
A 0.42$''\times0.35''$ of a 226 GHz (PA=-80$^\circ$)  
image showing the mini-spiral structure.  The image is primary beam corrected, thus, 
the noise at the edge of the beam is amplified. The proplyd 
candidates are located near the FWHM  of the primary beam  along the northeastern arm of the 
mini-spiral. 
{\it (b) Bottom}
Similar to (a) except that it is an 8 GHz image taken with the VLA in its A configuration (Yusef-Zadeh et al. 2016)
and is convolved  to the same resolution as  (a). The box in both figures shows
 the region where the brightest proplyds are concentrated.   The dark features along the 
Northern arms are likely to be radio dark clouds which 
are imprints of molecular gas against the  continuum emission (Yusef-Zadeh 2012).
}
\end{figure}


\begin{figure}
\center
\includegraphics[scale=0.6,angle=0]{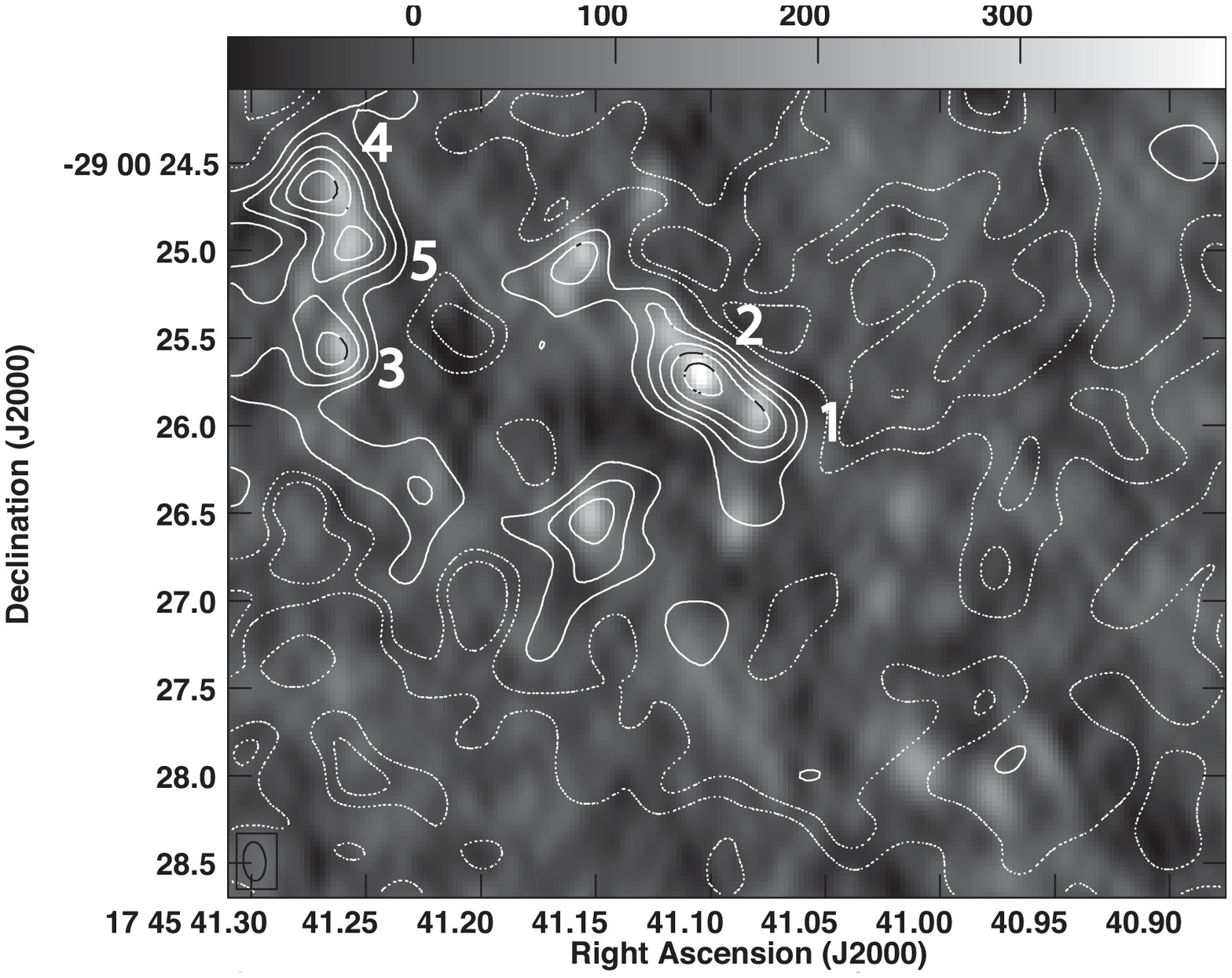}
\includegraphics[scale=0.6,angle=0]{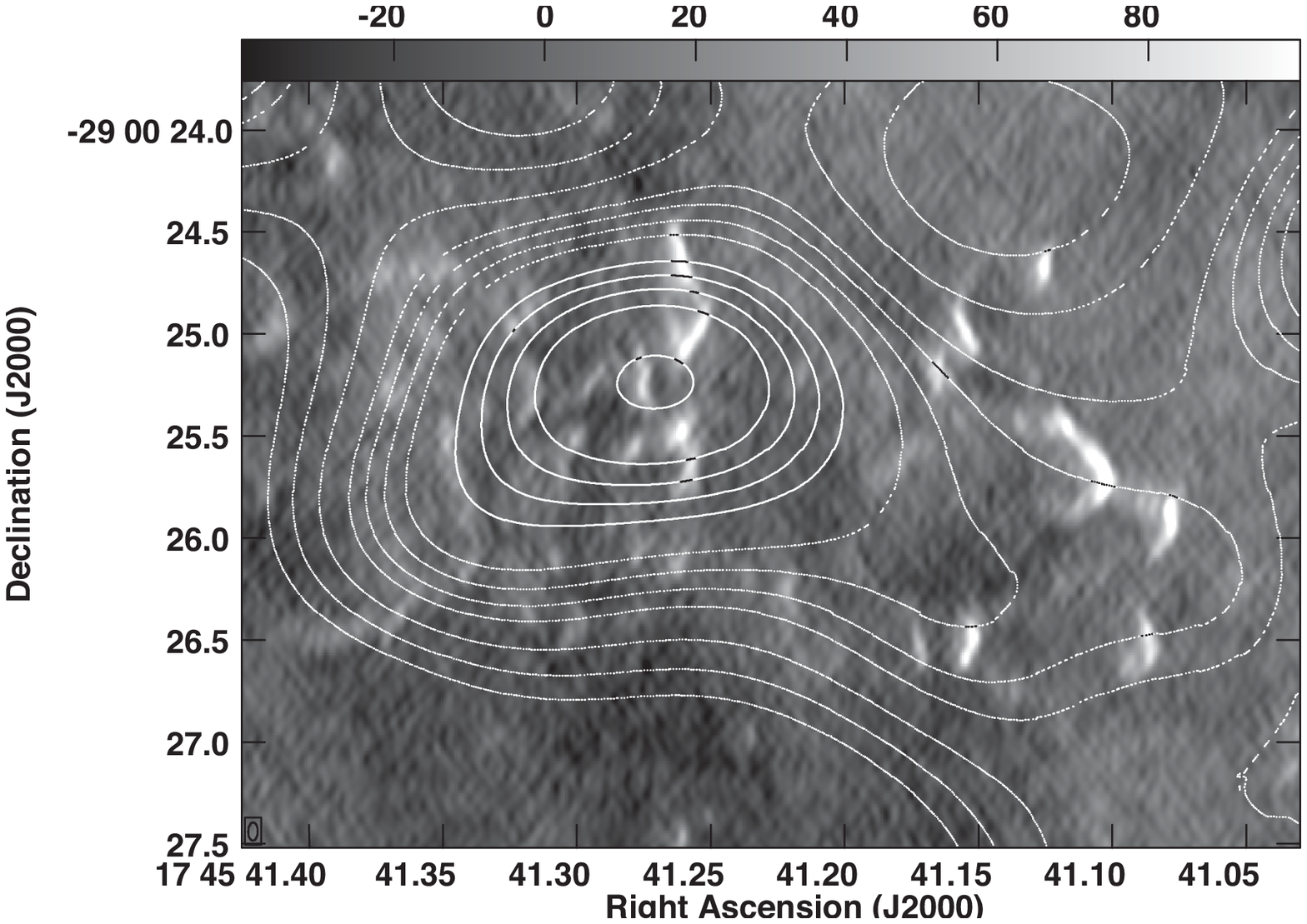}
\caption{
{\it (a) Top}
Contours of 226 GHz emission set at -0.4, -0.2, 0.2, 0.4, 0.6, 0.8, 1 mJy beam$^{-1}$
are superimposed on a grayscale 44 GHz image
with the same resolution as that  in Figure 1a.  No primary bean correction has been applied  
to 226 GHz data. 
{\it (b) Bottom}
Contours of 100 GHz emission set at (-2, -1.5, -1, -0.5, 0.5, 1, 1.5, 2, 3, 4) $\times 0.5$ mJy beam$^{-1}$
are superimposed on a grayscale 34 GHz image 
with spatial resolutions of 1.85$''\times 1.50''$(PA= -87$^\circ$.4)
and 0.088$\times0.046''$ (PA=-1$^\circ$.56), respectively. 
}
\end{figure}

\begin{figure}
\center
\includegraphics[scale=1,angle=0]{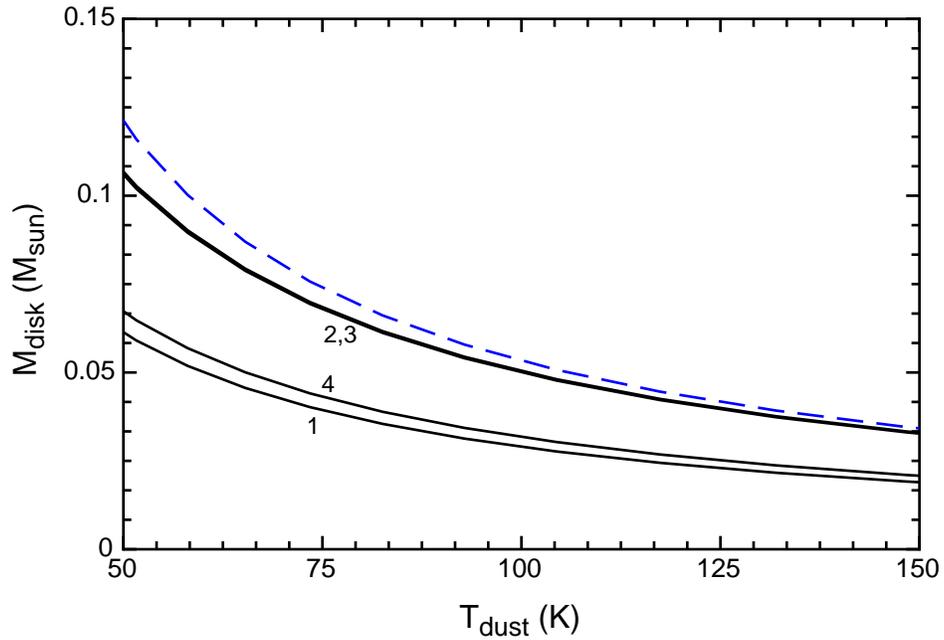}
\caption{
Estimated disk mass vs assumed dust temperature for the four sources detected in dust thermal continuum at 
226\,GHz after subtracting the bremsstrahlung contribution and assuming that the emission is optically thin.  
The blue dashed curve shows the disk mass for the brightest source corrected for finite optical depth assuming 
that it subtends $(300\,\mathrm{AU})^2$ in the
plane of the sky (see text).
}
\end{figure}
\end{document}